\documentclass[10pt,journal,final,twocolumn]{IEEEtran}

\usepackage{graphics}
\usepackage{subfigure}
\usepackage{epstopdf}
\usepackage{epsfig}
\usepackage[cmex10]{amsmath}
\usepackage{amssymb}
\usepackage{times}
\usepackage{ntheorem}
\usepackage{pifont}
\usepackage{stfloats}
\usepackage{bm}
\usepackage{color}
\usepackage{makecell}
\usepackage{array}
\usepackage{multirow}

\begin{document}
%
\title{Lightweight Deep Learning Based Channel Estimation for Extremely Large-Scale Massive MIMO Systems}

\author{\IEEEauthorblockN{
Shen~Gao, Peihao~Dong,~\IEEEmembership{Member,~IEEE}, Zhiwen~Pan,~\IEEEmembership{Member,~IEEE},
and Xiaohu~You,~\IEEEmembership{Fellow,~IEEE}
}

\thanks{

This work was supported in part by the National Science Foundation of China under Grant 62101253, the Natural Science Foundation of Jiangsu Province under Grant BK20210283, and the open research fund of National Mobile Communications Research Laboratory, Southeast University (2022D08). \emph{(Corresponding author: Peihao Dong.)}

S. Gao, Z. Pan, and X. You are with the National Mobile Communications Research Laboratory, Southeast University, Nanjing 211111, China, and also with the Purple Mountain Laboratories, Nanjing 211100, China. (e-mail: gaoshen@seu.edu.cn; pzw@seu.edu.cn; xhyu@seu.edu.cn).

P. Dong is with the College of Electronic and Information Engineering, Nanjing University of Aeronautics and Astronautics, Nanjing 211106, China, and also with the National Mobile Communications Research Laboratory, Southeast University, Nanjing 211111, China. (e-mail: phdong@nuaa.edu.cn).
}
}

\IEEEtitleabstractindextext{%
\begin{abstract}
Extremely large-scale massive multiple-input multiple-output (XL-MIMO) systems introduce the much higher channel dimensionality and incur the additional near-field propagation effect, aggravating the computation load and the difficulty to acquire the prior knowledge for channel estimation. In this article, an XL-MIMO channel network (XLCNet) is developed to estimate the high-dimensional channel, which is a universal solution for both the near-field users and far-field users with different channel statistics. Furthermore, a compressed XLCNet (C-XLCNet) is designed via weight pruning and quantization to accelerate the model inference as well as to facilitate the model storage and transmission. Simulation results show the performance superiority and universality of XLCNet. Compared to XLCNet, C-XLCNet incurs the limited performance loss while reducing the computational complexity and model size by about $10\times$ and $36\times$, respectively.
\end{abstract}

}

\maketitle

\IEEEdisplaynontitleabstractindextext

%
\IEEEpeerreviewmaketitle

\section{Introduction}

\IEEEPARstart{T}{he} success of massive multiple-input multiple-output (MIMO) in dramatically boosting spectral efficiency has been witnessed by both academia and industry \cite{L. Lu}. Nevertheless, the scale of the current massive MIMO array is inadequate to support ultra massive connections and services requiring ultrahigh data rate conceived in the sixth generation (6G) networks, and thus the extremely large-scale massive MIMO (XL-MIMO) was proposed \cite{W. Tong}--\cite{A. M. Elbir_a}.

To fully deliver the highlights of XL-MIMO, accurate channel state information (CSI) is necessary. In distinction to the conventional massive MIMO system where all users experience the far-field propagation effect, the extremely large number of antennas of XL-MIMO forces a part of users to suffer from the near-field effect and thus poses tough challenges for channel estimation. In \cite{Y. Han}, near-field channel estimation schemes were proposed from the subarray-wise and scatterer-wise perspectives for XL-MIMO systems. For the XL-MIMO system with grant-free access, the joint user activity and channel estimation algorithm was designed in \cite{H. Iimori} by exploiting both the subarray-wise sparsity and the activity sparsity. To reduce the pilot overhead for estimating the XL-MIMO downlink channel, the sparse representation of the near-field channel was uncovered in the polar-domain and then compressive sensing based algorithms were proposed in \cite{M. Cui}. In \cite{X. Wei}, a hybrid-field channel model inclusive of both far-field and near-field components was formulated and the channel estimation approach was developed accordingly, which was further modified by blending support detection in \cite{Z. Hu}. In \cite{J. Cao}, a sparse Bayesian learning aided off-grid channel estimation approach was proposed by using the improved sparsity of the near-field channel.

Since the channels are sparse in different domains, the existing works treat the near-field and far-field channel estimation tasks separately based on the exact knowledge of the user location and channel statistics, which are not always available in practice, especially in the high-speed mobile scenario. Then these approaches may suffer from the performance degradation due to the frequent mismatch between the prior knowledge and the dynamic propagation environment. Furthermore, the size of the antenna array can be augmented by even an order of magnitude from the current massive MIMO to XL-MIMO, which will aggravate the computation load for executing these approaches and thus leads to the intolerant service latency for 6G networks. It has been shown that deep learning (DL)-based massive MIMO channel estimation approaches are much less dependent on the prior knowledge and can invoke the parallel computing architecture to shorten the runtime by at least $10\times$ \cite{P. Dong}. Therefore, this article proposes to estimate the high-dimensional XL-MIMO channel via DL. The main novelty and contribution can be summarized as follows

\begin{itemize}[\IEEEsetlabelwidth{Z}]
\item[1)] An XL-MIMO channel network (XLCNet) tailored for the multi-user high-dimensional channel estimation is developed. Resorting to the properly designed model architecture and generation method of training data, XLCNet is applicable for both the near-field and far-field users with different channel statistics, independent of the prior knowledge.

\item[2)] To accelerate the model inference and to facilitate the model storage and transmission, a compressed XLCNet (C-XLCNet) is designed via weight pruning and quantization, which reduces the computational complexity and the model size by about $10\times$ and $36\times$, respectively, at the cost of very limited performance loss compared with XLCNet.
\end{itemize}

\emph{Notations}: In this article, we use upper and lower case boldface letters to denote matrices and vectors, respectively. $(\cdot)^T$, $\|\cdot\|_{F}$, and $\mathbb{E}\{\cdot\}$ represent the transpose, Frobenius norm, and expectation, respectively. $|a|$ denotes the modulus of a complex-valued number $a$. $\lfloor\cdot\rfloor$ denotes the floor function. $\mathcal{CN}(0,\sigma^2)$ represents a circular symmetric complex Gaussian distribution with variance $\sigma^2$.

\section{System Model}

\subsection{Signal Transmission Model}

\begin{figure}[t]
\centering
\includegraphics[width=2.8in]{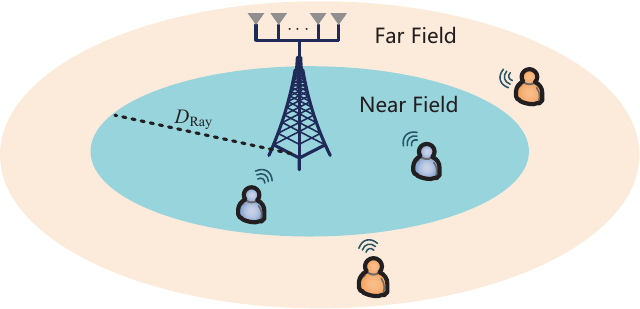}
\caption{A multi-user XL-MIMO system.}\label{System_model}
\end{figure}

As shown in Fig.~\ref{System_model}, consider a multi-user XL-MIMO system working in the time division duplex (TDD) mode. The reciprocal channel between the base station (BS) with extremely large number of antennas, i.e., $M$, and $K$ users can be acquired via uplink pilot signals, that is,
\begin{eqnarray}
\label{eqn_Y}
\mathbf{Y}=\mathbf{H}\mathbf{P}^{\frac{1}{2}}\mathbf{X}+\mathbf{Z},
\end{eqnarray}
where $\mathbf{H}\in \mathbb{C}^{M\times K}$, $\mathbf{P}=\textrm{diag}(P_1,\ldots,P_{K})$, $\mathbf{X}\in \mathbb{C}^{K\times K}$, and $\mathbf{Z}\in \mathbb{C}^{M\times K}$ denote the channel matrix, the transmit power matrix, the transmit pilot matrix, and the additive white Gaussian noise (AWGN) matrix, respectively. Without loss of generality, we assume $P_1=\ldots=P_{K}=P$ and $\mathbf{X}=\mathbf{I}_{K}$ for simplicity, which yields
\begin{eqnarray}
\label{eqn_Y_simp}
\mathbf{Y}=\sqrt{P}\mathbf{H}+\mathbf{Z}.
\end{eqnarray}

\subsection{Channel Model}

In the XL-MIMO system, the significantly increased array size introduces the near-field propagation effect of electromagnetic wave, which is different from the far-field effect considered in previous generations of MIMO systems \cite{M. Cui}, \cite{X. Wei}. In fact, the near-field and far-field are divided by the Rayleigh distance $D_{Ray}$. The electromagnetic wave propagation will be subject to the near-field effect and far-field effect when the distance between the transceiver, $D_{tr}$, satisfies $D_{tr}\leq D_{Ray}$ and $D_{tr}> D_{Ray}$, respectively. According to \cite{K. T. Selvan}, $D_{Ray}$ can be expressed as $D_{Ray}=\frac{2 D_{a}^2}{\lambda}$, where $D_{a}$ and $\lambda$ represent the array aperture and the carrier wavelength, respectively. Consider a uniform linear array (ULA) equipped at the BS with the space between adjacent antennas $d=\frac{\lambda}{2}$, we have $D_{Ray}=\frac{2 (M\frac{\lambda}{2})^2}{\lambda}=\frac{1}{2}M^2\lambda$, indicating the Rayleigh distance increases proportionally to the square of the number of BS antennas. For an XL-MIMO BS equipped with $200$ antennas and working at the frequency band of $30$GHz, $D_{Ray}$ can reach to $200$ meters. This non-trivial distance may cover a part of users and thus brings them the near-field propagation effect, as shown in Fig.~\ref{System_model}.

For the multi-user XL-MIMO system, the far-field channel model and the near-field channel model can be generally expressed in a unified form, that is
\begin{eqnarray}
\label{eqn_Hall}
\mathbf{H}=\mathbf{A}\mathbf{G},
\end{eqnarray}
where $\mathbf{A}\in \mathbb{C}^{M\times L}$ and $\mathbf{G}\in \mathbb{C}^{L\times K}$ denote the steering matrix and the path gain matrix composed of independent identically distributed (i.i.d.) $\mathcal{CN}(0,\sigma^2)$ elements, respectively, with $L$ representing the number of paths and $\sigma^2$ representing the average power gain. The main difference between the far-field channel model and the near-field channel model lies in $\mathbf{A}$. For the far-field channel model, we have $\mathbf{A}=\sqrt{\frac{M}{L}}[\mathbf{a}(\varphi_{1}),\ldots,\mathbf{a}(\varphi_{L})]$ and $\mathbf{a}(\varphi_{l})$ is further expressed as
\begin{eqnarray}
\label{eqn_a_varphi}
\mathbf{a}(\varphi_{l})=\frac{1}{\sqrt{M}}\left[1, e^{-j2\pi\frac{d}{\lambda}\sin\varphi_{l}},\ldots,e^{-j2\pi\frac{d}{\lambda}(M-1)\sin\varphi_{l}}\right]^T,
\end{eqnarray}
where $\varphi_{l}\in [-\frac{\pi}{2}, \frac{\pi}{2}]$ denotes the azimuth angle of arrival at the BS of the $l$th path. For the near-field case, $\mathbf{A}$ is parameterized by $\varphi_{l}$ and $r_{l}$ as $\mathbf{A}=\sqrt{\frac{M}{L}}[\mathbf{a}(\varphi_{1},r_{1}),\ldots,\mathbf{a}(\varphi_{L},r_{L})]$ and $\mathbf{a}(\varphi_{l},r_{l})$ is given by
\begin{eqnarray}
\label{eqn_a_varphi_r}
\mathbf{a}(\varphi_{l},r_{l})=\frac{1}{\sqrt{M}}\left[e^{-j\frac{2\pi}{\lambda}(r_{l,1}-r_{l})},\ldots,e^{-j\frac{2\pi}{\lambda}(r_{l,M}-r_{l})}\right]^T,
\end{eqnarray}
where $r_{l}$ denotes the distance between the $l$th scatterer and the center of the BS antenna array, $r_{l,m}=\sqrt{r_{l}^2+\delta_{m}^2 d^2-2r_{l}\delta_{m}d\sin\varphi_{l}}$ denotes the distance between the $l$th scatterer and the $m$th BS antenna with $\delta_{m}=\frac{2m-M-1}{2}$, $m=1,\ldots, M$.

\section{XLCNet Design}

In this section, XLCNet is developed to estimate the multi-user high-dimensional channel, $\mathbf{H}$. The preprocessing procedure is first introduced, followed by the elaboration on XLCNet architecture and training.

\subsection{Preprocessing}

Since the channels of $K$ users bear the same general structure according to (\ref{eqn_Hall}), we simplify the design by focusing on one user. Then (\ref{eqn_Y_simp}) reduces to the vector version as
\begin{eqnarray}
\label{eqn_y}
\mathbf{y}=\sqrt{P}\mathbf{h}+\mathbf{z},
\end{eqnarray}
based on which the least-square (LS) estimation of $\mathbf{h}$ is given by
\begin{eqnarray}
\label{eqn_h_ls}
\hat{\mathbf{h}}_{\text{LS}}=\frac{\mathbf{y}}{\sqrt{P}}=\mathbf{h}+\frac{\mathbf{z}}{\sqrt{P}}.
\end{eqnarray}
The coarse estimate $\hat{\mathbf{h}}_{\text{LS}}$ will be refined by XLCNet to yield a more accurate version.

\subsection{XLCNet Architecture}

In general, the architecture of XLCNet can be diverse. A direct way is using the fully-connected or one-dimensional (1D) convolutional structure to refine $\hat{\mathbf{h}}_{\text{LS}}$, which, however, results in the huge number of neural network weights or low training efficiency. Therefore, $\hat{\mathbf{h}}_{\text{LS}}$ will be reshaped into a matrix, $\hat{\bar{\mathbf{H}}}_{\text{LS}}$, and then processed by XLCNet composed of two-dimensional (2D) convolutional layers.\footnote{The 1D-convolution and 2D-convolution refer to operations on one feature map channel.} The residual structure is applied to achieve a better denoising effect. In more details, $\hat{\bar{\mathbf{H}}}_{\text{LS}}$ is input into XLCNet to approximate the true channel matrix, $\bar{\mathbf{H}}$, reshaped by $\mathbf{h}$. XLCNet includes two branches for data flow, one of which uses a series of 2D convolutional layers to extract the noise from $\hat{\bar{\mathbf{H}}}_{\text{LS}}$, and another one is a shortcut simply replicating $\hat{\bar{\mathbf{H}}}_{\text{LS}}$. By subtracting the extracted noise from $\hat{\bar{\mathbf{H}}}_{\text{LS}}$, a purer estimate, $\hat{\bar{\mathbf{H}}}$, is obtained. Nine zero padding (ZP) convolutional layers are used in the first branch for feature extraction. Each of the first eight layers applies $64$ $3 \times 3$ kernels, rectified linear unit (ReLU) activation function and batch normalization (BN) while the last layer applies $2$ $3 \times 3$ kernels and directly output the filtered results without the activation function and BN. In addition, it is preferable to stack the convolutional layers into one residual block rather than disperse these layers into multiple residual blocks since the former is more favorable to continuously extract the useful features \cite{S. Gao_b}, as shown in the following simulation results.

\subsection{XLCNet Training}

In the XL-MIMO cell, a part of users suffer from the near-field propagation effect while other users experience the far-field effect, depending on user locations. The situation of the propagation effect and channel characteristics may be frequently changed in the mobile communications. The BS should catch up with the dynamics to select the right channel estimation algorithm for each user, or the performance will be deteriorated due to the information mismatch. It is necessary to design a unified channel estimation framework applicable for users experiencing different propagation effects and channel statistics, in order to get rid of the dependence on the dynamic information.

Motivated by the goal, XLCNet is trained based on the data generated by the hybrid-field channel model \cite{X. Wei}. According to (\ref{eqn_Hall}), $\mathbf{h}$ is expressed as
\begin{eqnarray}
\label{eqn_h_hyb}
\mathbf{h}=\mathbf{A}_{\textrm{hy}}\mathbf{g},
\end{eqnarray}
where the path gain vector $\mathbf{g}\in \mathbb{C}^{L\times 1}$ is composed of i.i.d. $\mathcal{CN}(0,\sigma^2)$ elements. $\mathbf{A}_{\textrm{hy}}=\sqrt{\frac{M}{L}}[\mathbf{a}(\varphi_{1}),\ldots,\mathbf{a}(\varphi_{L_{0}}), \\ \mathbf{a}(\varphi_{L_{0}+1},r_{L_{0}+1}), \ldots,\mathbf{a}(\varphi_{L},r_{L})]$ indicates that $\mathbf{h}$ consists of $L_{0}$ far-field paths and $L-L_{0}$ near-field paths, that is
\begin{eqnarray}
\label{eqn_h_hyb_sum}
\mathbf{h}=\sqrt{\frac{M}{L}}\left[\sum_{l=1}^{L_{0}}g_{l}\mathbf{a}(\varphi_{l})+\sum_{l=L_{0}+1}^{L}g_{l}\mathbf{a}(\varphi_{l},r_{l})\right].
\end{eqnarray}
Selecting the appropriate values of $L_{0}$ and $L$ can help XLCNet improve the training efficiency and the generalization ability in scenarios with different propagation effects and numbers of paths. Specifically, $L_{0}$ determines the proportions of far-field paths and near-field paths. From (\ref{eqn_a_varphi}) and (\ref{eqn_a_varphi_r}), the near-field component has the more complicated structure and thus a larger proportion is allocated to facilitate the learning of the inherent structure. So the value of $L_{0}$ is taken from the set $\{1,\ldots,\lfloor\frac{L}{2}\rfloor\}$. The another hyperparameter, $L$, denotes the total number of paths for training data generation. It has been shown that the CNN performs well in the testing scenario with a smaller value of $L$ than that used in the training stage and vice versa \cite{P. Dong}. So $L$ needs to be set larger than the possible numbers of paths at the working frequency band. Given $L_{0}$ and $L$, $\mathbf{h}$ formulated in (\ref{eqn_h_hyb_sum}) is used to generate training dataset, $\mathcal{D}_{\textrm{tr}}$, based on which XLCNet is trained to minimize the mean-squared error (MSE) loss,
\begin{eqnarray}
\label{eqn_Loss}
\mathcal{L}=\frac{1}{N_{\textrm{tr}}}\sum_{n=1}^{N_{\textrm{tr}}}\|\bar{\mathbf{H}}^{(n)}-\hat{\bar{\mathbf{H}}}^{(n)}\|_F^2,
\end{eqnarray}
where $N_{\textrm{tr}}$ denotes the number of training samples and the superscript $(n)$ indicates the $n$th sample.

\emph{Remark 1 (Why XLCNet Applicable for Both Near-field and Far-field Users?):} The reason is mainly three-fold. 1) The far-field channel model and the near-field channel model share the unified structure given by (\ref{eqn_Hall}) while just differing in the detailed form of the steering matrix $\mathbf{A}$. 2) The training data of XLCNet are generated based on the hybrid-field channel model given by (\ref{eqn_h_hyb_sum}) so that both the far-field and near-field components can be incorporated. 3) The input vector is reshaped into the matrix and sufficient number of 2D convolutional layers are applied in XLCNet for the better feature extraction. The three reasons facilitate the learning of a universal feature space for both the near-field and far-field channels.

\section{Lightweight Design of XLCNet}

In this section, we focus on the lightweight design of XLCNet to reduce the complexity as well as to facilitate the model storage and transmission. XLCNet will be made sparse by weight pruning and then the remaining weights are quantized, yielding C-XLCNet. The computational complexity of C-XLCNet is finally analyzed to shed light on the effect of model compression.

\subsection{Weight Pruning}

Weight pruning aims to cut off the dispensable weights while maintaining a satisfactory accuracy. Following the magnitude pruning criterion, the weights with absolute values less than the threshold are regarded as trivial contributors to the performance and will be removed from XLCNet, after which the reserved weights are fine-tuned to compensate the performance loss.\footnote{Although reducing the number of network layers is a simple way to decrease the model size and complexity, it will lead to the insufficient representation of the desired feature space and thus degrades the estimation accuracy according to the simulation trails. Instead, it is better to remove the trivial weights of each layer to achieve the same goal at the cost of very limited performance loss.}

In the pruning procedure, the pruning threshold needs to be determined first. Denote $\boldsymbol{\Theta}=\{\mathbf{W}_{1},\ldots,\mathbf{W}_{C}\}$ as the weight set of XLCNet, where $C$ denotes the total number of convolutional layers and the four-dimensional tensor, $\mathbf{W}_{c}$, denotes the kernel weights used by the $c$th layer, $c=1,\ldots,C$. Sort absolute values of all weights in $\boldsymbol{\Theta}$ in an ascending order to yield a vector, $\boldsymbol{\theta}$, and select the $\lfloor\kappa N_{\textrm{w}}\rfloor$th element of $\boldsymbol{\theta}$, i.e., $\theta_{\lfloor\kappa N_{\textrm{w}}\rfloor}$, as the threshold, where $N_{\textrm{w}}$ denotes the total number of weights in $\boldsymbol{\Theta}$ and the pruning ratio, $\kappa$, indicates the proportion of weights removed. Then each element of $\boldsymbol{\Theta}$ is processed as
\begin{eqnarray}
\label{eqn_Prun}
[\mathbf{W}_{c}]_{i}=
\begin{cases}
	[\mathbf{W}_{c}]_{i}, & \left| [\mathbf{W}_{c}]_{i}\right| \geq \theta_{\lfloor\kappa N_{\textrm{w}}\rfloor}\\
	0, & \textrm{otherwise},
\end{cases},
\end{eqnarray}
where $[\mathbf{W}_{c}]_{i}$ denotes the $i$th element of $\mathbf{W}_{c}$ in the certain index way for $c=1,\ldots,C$ and $i=1,\ldots,N_{c}$ with $N_{c}$ representing the total number of elements of $\mathbf{W}_{c}$. Note that setting a weight as $0$ based on (\ref{eqn_Prun}) means this weight becomes invalid and will not cause computational cost in the subsequent model inference. Finally, the sparse XLCNet after pruning will be fine-tuned within a certain number of epochs to achieve the estimation accuracy closer to the original XLCNet, still based on the training set, $\mathcal{D}_{\textrm{tr}}$, and the loss function, $\mathcal{L}$, in (\ref{eqn_Loss}).

\subsection{Weight Quantization}

Although most weights are removed in the pruning procedure, the remaining weights are still $32$-bit floating-point numbers occupying the excessive computation resource and storage space, which can be resolved by weight quantization.

The post-training quantization is considered herein, which is a push-button way without need of re-training. Denoting $b$ as the bit-width, the general uniform affine quantization is applied to map each reserved weight after pruning to the unsigned integer set $\{0,\ldots,2^{b}-1\}$ \cite{M. Nagel}, that is,
\begin{eqnarray}
\label{eqn_Gen_Quan}
\mathbb{Q}\left([\mathbf{W}_{c}]_{i}\right)=\textrm{clamp}\left(\textrm{round}\left(\frac{[\mathbf{W}_{c}]_{i}}{S}\right)+Z;0,2^{b}-1\right),
\end{eqnarray}
where $\textrm{round}(\cdot)$ denotes the round-to-nearest operation, $\textrm{clamp}(\cdot;\cdot,\cdot)$ is defined as
\begin{eqnarray}
\label{eqn_clamp}
\textrm{clamp}(x;\mu_1,\mu_2)=
\begin{cases}
	\mu_1, & x<\mu_1,\\
	x, & \mu_1 \leq x \leq \mu_2,\\
    \mu_2, & x>\mu_2,
\end{cases}.
\end{eqnarray}
For (\ref{eqn_Gen_Quan}), $S$ is a scaling factor and $Z$ is an integer zero point representing the quantization of the real zero, which are respectively expressed as
\begin{eqnarray}
\label{eqn_ScaleFactor}
S=\frac{\max(\mathbf{W}_{c})-\min(\mathbf{W}_{c})}{2^{b}-1},
\end{eqnarray}
\begin{eqnarray}
\label{eqn_ZeroPoint}
Z=-\textrm{round}\left(\frac{(2^{b}-1)\min(\mathbf{W}_{c})}{\max(\mathbf{W}_{c})-\min(\mathbf{W}_{c})}\right),
\end{eqnarray}
where $\max(\mathbf{W}_{c})$ and $\min(\mathbf{W}_{c})$ denote the maximum value and the minimum value among the reserved weights of the $c$th convolutional layer. After weight pruning and quantization, the lightweight C-XLCNet is obtained with a model compression ratio
\begin{eqnarray}
\label{eqn_CompreRatio}
\gamma=\frac{32}{b(1-\kappa)},
\end{eqnarray}
where the value of $b$ is generally set no more than $8$, indicating that C-XLCNet has a significantly reduced size facilitating the model storage and transmission.

\begin{figure}[t]
\centering
\includegraphics[width=3.2in]{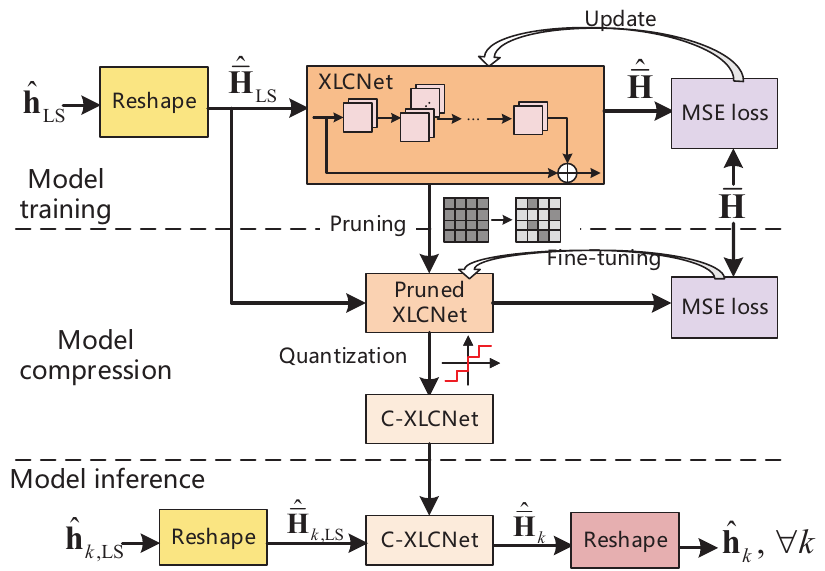}
\caption{Workflow of the lightweight DL based channel estimation framework.}\label{XLCNet}
\end{figure}

The workflow of the proposed framework including model training, model compression, and model inference is summarized in Fig.~\ref{XLCNet}.

\subsection{Complexity Analysis}

In this subsection, the computational complexity of C-XLCNet in the model inference stage is analyzed. The complexity of floating-point operations (FLOPs) is selected as the metric.

For the original XLCNet, the complexity is given by
\begin{eqnarray}
\label{eqn_Complexity}
\mathcal{C}_{\textrm{XLCNet}}\sim\mathcal{O}\left(\sum_{c=1}^{C}D_{c,1}D_{c,2}F_{c}^2 N_{c-1} N_{c}\right),
\end{eqnarray}
where for the $c$th convolutional layer, $D_{c,1}$ and $D_{c,2}$ denote the length and width of output feature maps, $F_{c}$ denotes the side length of the kernel, $N_{c-1}$ and $N_{c}$ denote the numbers of input and output feature maps. Since ZP convolution is applied for $c=1,\ldots,C$, $D_{c,1}D_{c,2}=M$ holds, yielding a compact form as
\begin{eqnarray}
\label{eqn_Complexity_Compact}
\mathcal{C}_{\textrm{XLCNet}}\sim\mathcal{O}\left(M\sum_{c=1}^{C}F_{c}^2 N_{c-1} N_{c}\right)=\mathcal{O}\left(M N_{\textrm{w}}\right),
\end{eqnarray}
which reveals that the complexity is mainly determined by the array size at the BS and the number of model weights. Then the complexity of C-XLCNet with the number of weights $(1-\kappa)N_{\textrm{w}}$ is given by
\begin{eqnarray}
\label{eqn_Complexity_C-XLCNet}
\mathcal{C}_{\textrm{C-XLCNet}}\sim\mathcal{O}\left(M (1-\kappa)N_{\textrm{w}}\right),
\end{eqnarray}
which is lower than the complexity of the original XLCNet by the factor of $\frac{1}{1-\kappa}$.

\section{Simulation Results}

In this section, simulation results are present to validate the effectiveness of XLCNet and C-XLCNet. The key simulation parameters are set as the number of BS antennas $M=256$, the wavelength $\lambda=0.01$ meters, the average path gain $\sigma^2=1$, $\varphi_{l}\in \mathcal{U}(-\frac{\pi}{2}, \frac{\pi}{2})$, and $r_{l}\in \mathcal{U}(10, 80)$ meters. For XLCNet, the training set and the validation set, generated based on the hybrid-field channel model in (\ref{eqn_h_hyb_sum}) with $L=6$ and $L_0=1$, contain $90,000$ and $10,000$ samples, respectively, and the testing set, generated based on the near-field or far-field channel model, contains $2,000$ samples. Adam is applied as the optimizer and the batch size is set as $128$. The model training lasts $200$ epochs with the learning rate $10^{-3}$ and the fine-tuning after pruning lasts $50$ epochs. The architecture parameters of XLCNet are detailed in Section III.B. The signal-to-noise ratio (SNR) is defined as $\mathrm{SNR}=10\lg\frac{P}{\sigma_{0}^2} \mathrm{(dB)}$ with the average noise power $\sigma_{0}^2$ normalized to $1$. The normalized MSE (NMSE), $\mathbb{E}\{\|\bar{\mathbf{H}}-\hat{\bar{\mathbf{H}}}\|_F^2/\|\bar{\mathbf{H}}\|_F^2\}$, is used to measure the channel estimation performance. The baseline schemes for performance comparison include the LS estimator, linear minimum MSE (LMMSE) estimator, hybrid-field orthogonal matching pursuit (HOMP) \cite{X. Wei}, and a variant version of multiple residual dense network \cite{Y. Jin}, named V-MRDN, that is tailored for both near-field and far-field users. The simulations are conducted on Intel(R) Core(TM) i7-9700 central processing unit (CPU) and NVIDIA GeForce RTX 2060 graphic processing unit (GPU).

\begin{figure}[t!]
\centering
\includegraphics[width=3.6in]{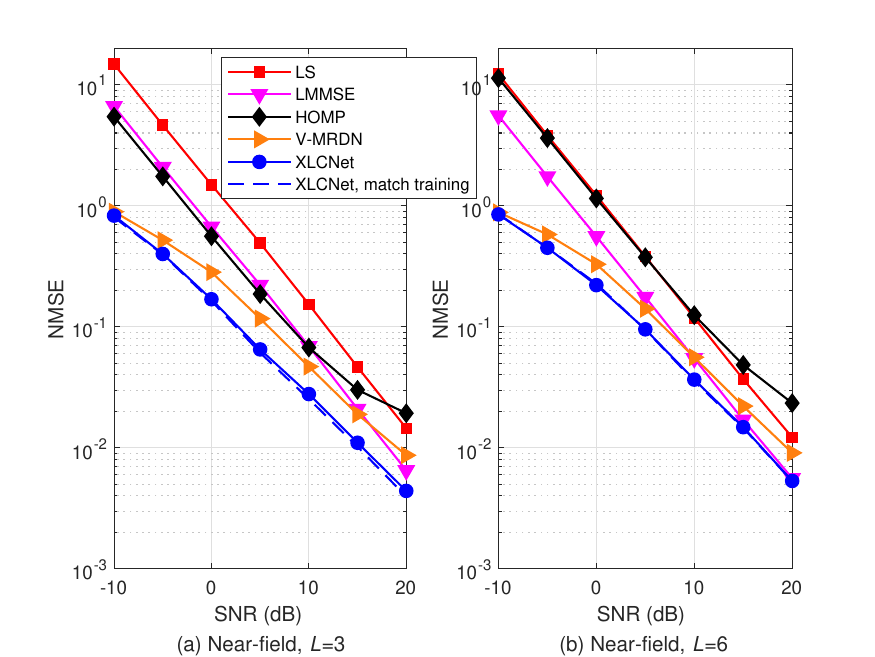}
\caption{NMSE versus SNR for the near-field user.}\label{NMSE_vs_snr_near}
\end{figure}

Fig.~\ref{NMSE_vs_snr_near} shows the NMSE performance versus SNR for the near-field user, where two subfigures corresponding to $L=3$ and $L=6$ share the legend. For the subfigures of Figs. 3 and 4, ``XLCNet" and ``V-MRDN" are trained using the same dataset with $L_{0}=1$ far-field path and $L-L_{0}=5$ near-field paths. In contrast, ``XLCNet, match training" is trained using different datasets with the corresponding propagation effects and numbers of paths indicated by the respective subfigure captions. From Fig.~\ref{NMSE_vs_snr_near}, the proposed XLCNet outperforms the baseline schemes, especially for $L=3$. In addition, XLCNet does not rely on the knowledge of channel statistics and can invoke the parallel computing to significantly accelerate the model inference, which are another two advantages compared with the LMMSE estimator and HOMP. The performance superiority of XLCNet over V-MRDN reveals that the network structure with fewer residual blocks and more convolutional layers inside is preferable for the considered channel estimation task. Match training means that the training set of XLCNet is generated with the same propagation effect and the same number of paths as the testing set. Although XLCNet is trained with $L=6$ and $L_0=1$, it achieves the almost same accuracy as the match training case, revealing the robustness when facing with different numbers of paths.

\begin{figure}[t!]
	\centering
	\includegraphics[width=3.6in]{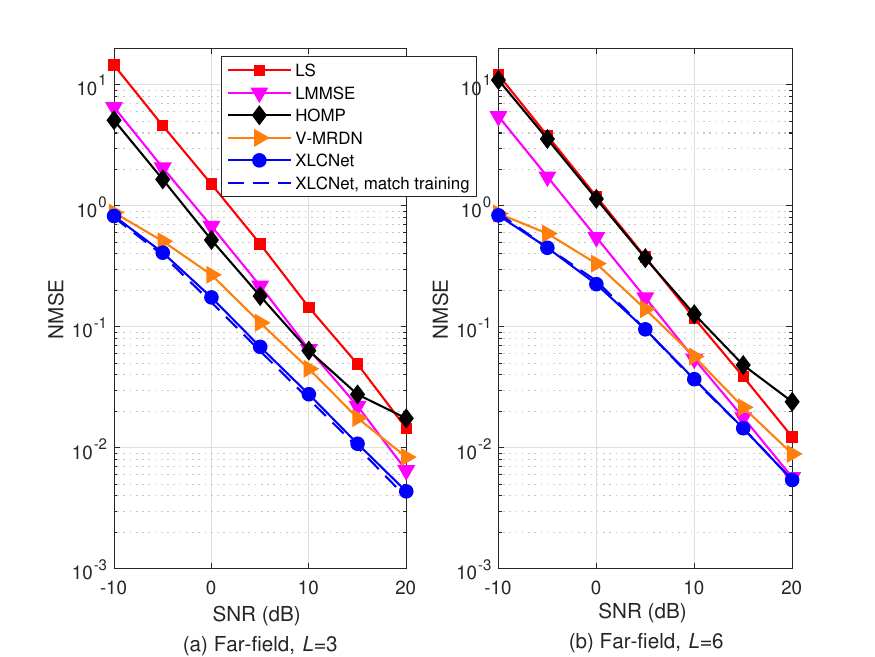}
	\caption{NMSE versus SNR for the far-field user.}\label{NMSE_vs_snr_far}
\end{figure}

Fig.~\ref{NMSE_vs_snr_far} shows the counterpart for the far-field user, where the performance gain of XLCNet remains almost unchanged. In addition, the performance of XLCNet still tightly fits the match training cases, validating that the proposed XLCNet is a universal solution applicable to users both in the near field and far field with different channel statistics. Furthermore, by comparing Fig.~\ref{NMSE_vs_snr_near} and Fig.~\ref{NMSE_vs_snr_far}, XLCNet provides the almost equal estimation accuracy for the near-field user and far-field user with the same number of paths.

\begin{figure}[t!]
	\centering
	\includegraphics[width=3.6in]{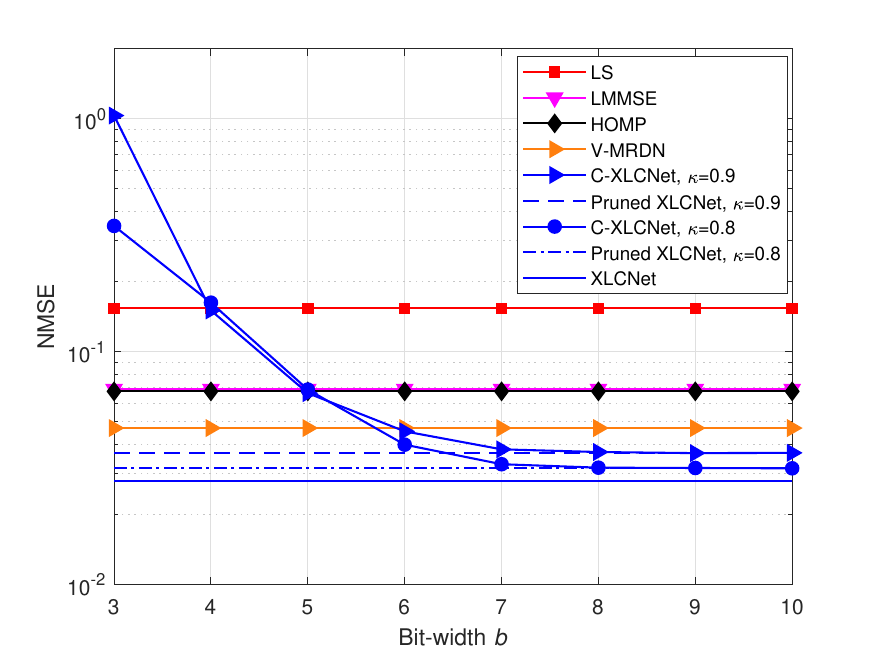}
	\caption{NMSE versus bit-width $b$ for the near-field user with $L=3$.}\label{NMSE_vs_bit_near}
\end{figure}

Considering that XLCNet performs consistently for near-field and far-field users, the performance of C-XLCNet versus the bit-width, $b$, is studied under the representative near-field case with $L=3$ and $\mathrm{SNR}=10$ dB in Fig.~\ref{NMSE_vs_bit_near}. Two values of the pruning ratio, $\kappa=0.8$ and $\kappa=0.9$, are considered to show the impact of weight pruning on the estimation accuracy. From Fig.~\ref{NMSE_vs_bit_near}, the pruned XLCNet with $\kappa=0.8$ performs very close to XLCNet and the performance loss is still limited even if $\kappa$ is increased to $0.9$. As expected, the performance of C-XLCNet decreases with $b$ and converges to the performance of pruned XLCNet with the corresponding $\kappa$. When $b\geq 6$, C-XLCNet starts to outperform the baseline schemes. As $b$ further increases to $8$, the performance of C-XLCNet becomes almost the same as the corresponding convergence value and close to the performance of XLCNet. Finally, the number of weights and run time of the proposed framework are listed in Table~\ref{weights_time}. From the table, weight pruning significantly reduces the number of weights and run time. For C-XLCNet, the additional weight quantization operation further shortens the run time remarkably. According to Table~\ref{weights_time} and (\ref{eqn_CompreRatio}), the model size of C-XLCNet is reduced by about $19\times$ and $36\times$ for $\kappa=0.8$ and $\kappa=0.9$, respectively.

\begin{table}[t]
	\centering
	\caption{Number of Weights and Run Time of the Proposed Framework}
    \label{weights_time}
	\begin{tabular}{c|c|c}
		\hline
		~ & Weights & \makecell{Run time (ms)} \\
		\hline
		XLCNet &263K &0.2 \\
		\hline
		\makecell{Pruned XLCNet, $\kappa=0.8$} &55K &0.13  \\
		\hline
		\makecell{C-XLCNet, $\kappa=0.8$, $b=8$} &55K &0.087  \\
		\hline
		\makecell{Pruned XLCNet, $\kappa=0.9$} &29K &0.12  \\
		\hline
		\makecell{C-XLCNet, $\kappa=0.9$, $b=8$} &29K &0.078  \\ \hline
	\end{tabular}
\end{table}

\section{Conclusion}

In this article, a universal channel estimation framework, named XLCNet, for the multi-user XL-MIMO system is proposed to mitigate the aggravated problems in terms of computation load and dependence on the channel prior knowledge. C-XLCNet is further designed to guarantee the fast model inference and the small model size. C-XLCNet reduces the computational complexity and the model size by about $10\times$ and $36\times$, respectively, at the cost of limited performance loss compared with XLCNet.



\ifCLASSOPTIONcaptionsoff
  \newpage
\fi


\begin{thebibliography}{99}
\bibitem{L. Lu}L. Lu, G. Y. Li, L. A. Swindlehurst, A. Ashikhmin, and R. Zhang, ``An overview of massive MIMO: benefits and challenges," \emph{IEEE J. Sel. Topics in Signal Process.}, vol. 8, no. 5, pp. 742--758, Oct. 2014.
\bibitem{W. Tong}W. Tong and P. Zhu, \emph{6G, the Next Horizon: From Connected People and Things to Connected Intelligence}. Cambridge University Press, 2021.
\bibitem{E. D. Carvalho}E. D. Carvalho, A. Ali, A. Amiri, M. Angjelichinoski, and R. W. Heath, Jr., ``Non-stationarities in extra-large-scale massive MIMO," \emph{IEEE Wireless Commun.}, vol. 27, no. 4, pp. 74--80, Aug. 2020.
\bibitem{A. M. Elbir_a}A. M. Elbir, K. V. Mishra and S. Chatzinotas, ``Terahertz-band joint ultra-massive MIMO radar-communications: Model-based and model-free hybrid beamforming," \emph{IEEE J. Sel. Topics in Signal Process.}, vol. 15, no. 6, pp. 1468--1483, Nov. 2021.
\bibitem{Y. Han}Y. Han, S. Jin, C.-K.Wen, and X. Ma, ``Channel estimation for extremely large-scale massive MIMO systems," \emph{IEEE Wireless Commun. Lett.}, vol. 9, no. 5, pp. 633--637, May 2020.
\bibitem{H. Iimori}H. Iimori \emph{et al.}, ``Joint activity and channel estimation for extra-large MIMO systems," \emph{IEEE Trans. Wireless Commun.}, to be published.
\bibitem{M. Cui}M. Cui and L. Dai, ``Channel estimation for extremely large scale MIMO: Far field or near field," \emph{IEEE Trans. Commun.}, vol. 70, no. 4, pp. 2663--2677, April 2022.
\bibitem{X. Wei}X. Wei and L. Dai, ``Channel estimation for extremely large-scale massive MIMO: Far-field, near-field, or hybrid-field," \emph{IEEE Commun. Lett.}, vol. 26, no. 1, pp. 177--181, Jan. 2022.
\bibitem{Z. Hu}Z. Hu, C. Chen, Y. Jin, L. Zhou and Q. Wei, ``Hybrid-field channel estimation for extremely large-scale massive MIMO system," \emph{IEEE Commun. Lett.}, vol. 27, no. 1, pp. 303--307, Jan. 2023.
\bibitem{J. Cao}J. Cao, J. Du, M. Han, J. Liu, X. Li and D. B. da Costa, ``Efficient sparse Bayesian channel estimation for near-field ultra-scale massive MIMO systems," \emph{IEEE Wireless Commun. Lett.}, to be published.
\bibitem{P. Dong}P. Dong, H. Zhang, G. Y. Li, I. Gaspar, and N. NaderiAlizadeh, ``Deep CNN-based channel estimation for mmWave massive MIMO systems," \emph{IEEE J. Sel. Topics Signal Process.}, vol. 13, no. 5, pp. 989$-$1000, Sep. 2019.
\bibitem{K. T. Selvan}K. T. Selvan and R. Janaswamy, ``Fraunhofer and Fresnel distances: Unified derivation for aperture antennas," \emph{IEEE Antennas Propag. Mag.}, vol. 59, no. 4, pp. 12--15, Aug. 2017.
\bibitem{S. Gao_b}S. Gao, P. Dong, Z. Pan, and G. Y. Li, ``Deep multi-stage CSI acquisition for reconfigurable intelligent surface aided MIMO systems," \emph{IEEE Commun. Lett.}, vol. 25, no. 6, pp. 2024--2028, June 2021.
\bibitem{M. Nagel}M. Nagel \emph{et al.}, ``A white paper on neural network quantization," \emph{arXiv preprint arXiv: 2106.0829}, 2021.
\bibitem{Y. Jin}Y. Jin \emph{et al.}, ``Multiple residual dense networks for reconfigurable intelligent surfaces cascaded channel estimation," \emph{IEEE Trans. Veh. Technol.}, vol. 71, no. 2, pp. 2134--2139, Feb. 2022.



\end{thebibliography}
\end{document}